\documentclass[12pt]{article}

\title{Semiclassical description of anisotropic magnets for spin $ S=1$}
\author{ Khikmat Kh. Muminov; Yousef Yousefi \\
Physical-Technical Institute named after S.U.Umarov\\
 Academy of Sciences of Republic of Tajikistan\\
Aini Ave 299/1, Dushanbe, Tajikistan\\
e-mail:  Khikmat@inbox.ru; yousof54@yahoo.com}
\date{}
\begin{document}
\maketitle
\begin{abstract}
In this paper, nonlinear equations describing one-dimensional non-Heisenberg ferromagnetic model are studied by use of generalized coherent states in a real parameterization. Also dissipative spin wave equation for dipole and quadruple branches is obtained if there is a small linear excitation from the ground state.
\end{abstract}
\section{Introduction}

In the past decades, magnets with spin value $s=\frac{1}{2}$  have been studied completely. There are dipoles, quadruples and higher order branches that affect the behavior of magnet crystal. However, the only necessary tool for describing the behavior of this kind of magnets is dipole branch effect and the order branches are not necessary. This results in linear approximation for describing the magnet behavior. 

Indeed, only dipole branch effect has been used for describing magnets with spin value $s\geq1$, and the effect of quadrupole and higher order branches have been ignored. Recently, however, due to the new developments in mathematics and technology and also due to the great potential of quadrupole branch in description of nano particles, its important role cannot be ignored.[1]. 

Using the effects of both dipole and quadrupole branches results in a nonlinear approximation. The use of higher order multipole effects yields more accurate approximatioms which demand more complicated equations. In this paper, only the effect of quadruple branch for Hamiltonians described by equation (1) is considered. Study of isotropic and anisotropic spin Hamiltonian with non-Heisenberg terms are complicated due to quadruple excitation dynamics [2,3,4]. Anti ferromagnetic property of this excitation in states near the ground, proves the existence of it. The effect of this calculation has been studied by Dzyaloshinskii [5]. The results obtained through the quadrupole excitation in nano particles $Fe_8$  and $Mn_{12}$ are more in line with numerical calculations and laboratory results [6,7].   

In classical physics term, the number of parameters required for a full macroscopic description of the magnet behavior is equal to $4s$  , where s is the spin value. Also real-parameterized coherent states based on related group is used to obtain classical equation of motion and to describe multipole dynamics [8,9].  Here Heisenberg ferromagnets with anisotropic term as described by equation (1) are considered:

\begin{eqnarray}
\hat H=-J\sum_i(\vec{ \hat S_i} \vec{\hat S_{i+1}}+\delta\hat S_i^z \hat S_i^z)
\end{eqnarray}

Here $ \hat S_i^x , \hat S_i^y ,\hat S_i^z$   are the spin operators acting at a site i, and $\delta$  is the anisotropy coefficient. This Hamiltonian is related to a one-dimensional ferromagnetic spin chains and the coefficient J is positive.

In order to calculate the effect of quadrupole excitation, first, the classical equivalent of Hamiltonian (1) is obtained and then by analyzing such equation for small linear excitation from the ground states, the spin wave solution is found. this process requires following steps:

1-	Obtaining coherent states for spin s=1 which are coherent states of SU(3) group. 

2-	Calculating the average values of spin operator.

3-	Obtaining classical spin Hamiltonian equation using previously calculated values.

4-	Computing Lagrangian equation by use of Feynman path integral over coherent states and then computing classical equations of motion.

5-	 For finding nonlinear equations of magnet behavior, it is necessary to substitute resulted Hamiltonian in classical equations of motion. Solutions of these nonlinear equations result in soliton description of magnet that is not interested here.

6-	Calculating ground states of magnet and then linearizing the nonlinear equations around the ground states for small excitation.

7-	At the end, calculating spin wave equation and dispersion equation.

In what fallows, the mathematical descriptions of the above steps are presented:

\section{Theory and Calculation}

In quantum mechanics, coherent states are special kind of quantum states that their dynamics are very similar to their corresponding classical system. These states are obtained by act of weil-Heisenberg group operator on vacuum state. Vacuum state of SU(3) group is $(1,0,0)^T$  and coherent state is introduced as [10]:
\begin{eqnarray}
|\psi \rangle &=& D^{\frac{1}{2}}(\theta, \phi)e^{-i\gamma \hat S^z} e^{2ig\hat Q^{xy}}|0\rangle \nonumber\\
&=& C_0|0\rangle +C_1|1\rangle+C_2|2\rangle
\end{eqnarray}

where $D^{1/2}(\theta, \phi)$ is wigner function and $Q^{xy}$  is quadruple moment which is written in the following form:

\begin{eqnarray}
\hat Q^{xy}=\frac{i}{2} \left(
\begin{array}{ccc}
0 & 0 &1 \\
0 & 0 & 0 \\
-1 & 0 & 0 
\end{array}
\right) \;
\end{eqnarray}

Coefficients $C_0$ to $ C_2$ computed from these equations: 
\begin{eqnarray}
C_0 &=& e^{i\phi}(e^{-i\gamma}sin^2({\theta/2})cosg+e^{i\gamma}cos^2({\theta/2})sing) \nonumber\\
C_1 &=&\frac{ sin{\theta}}{\sqrt{2}} (e^{-i\gamma}cosg-e^{i\gamma}sing) \nonumber\\
C_2 &=& e^{-i\phi}(e^{-i\gamma}cos^2({\theta/2})cosg+e^{i\gamma}sin^2({\theta/2})sing)
\end{eqnarray}

Two angles, $\theta$  and $\phi$ , determine the direction of classical spin vector in spherical coordinate system. The angle $\gamma$  determines the direction of quadruple moment around the spin vector and parameter g shows change of magnitude of spin vector. 

In order to obtain the classical equivalent of Hamiltonian (1), the classical equivalent of spin vector and their corresponding products should be computed. So consider:

\begin{eqnarray}
\vec S=\langle\psi|\vec {\hat S}|\psi\rangle
\end{eqnarray}

 as classical spin vector, and also consider: 

\begin{eqnarray}
Q^{ij}=\frac{1}{2}(\hat S_i \hat S_j+\hat S_j \hat S_i -\frac{4}{3}\delta_{ij}I)
\end{eqnarray}

components of quadruple moment. Spin operators can be commute in different lattices; so

\begin{eqnarray}
\langle\psi|\hat S_{n}^i \hat S_{n+1}^j|\psi\rangle= \langle\psi | \hat S_n^i |\psi \rangle  \langle\psi | \hat S_{n+1}^j |\psi \rangle 
\end{eqnarray}

where $ |\psi \rangle=|\psi\rangle_n | \psi\rangle_{n+1}$.

The average spin values in SU(3) group are defined as [11]:
\begin{eqnarray}
S^+ &=& e^{i\phi}cos(2g)sin\theta  \nonumber\\
S^- &=& e^{-i\phi} cos(2g)sin\theta \nonumber\\
S^z &=& cos(2g) cos\theta \nonumber\\
S^2&=& cos^2 (2g)
\end{eqnarray}

classical Hamiltonian can be obtained from the average calculation of Hamiltonian (1) over coherent states. The classical continuous limit of Hamiltonian in SU(3) group is:

\begin{eqnarray}
H_{cl}&=& -J\int \frac{dx}{a_0}(cos^2( 2g)+\frac{\delta}{2}(cos^2\theta+sin(2g)cos(2\gamma) sin^2\theta) \nonumber\\
&  &-\frac{a_0^2}{2}((\theta_x^2+\phi_x^2 sin^2\theta)cos^2( 2g)+4g_x^2 sin^2( 2g)))
\end{eqnarray}

The above classical Hamiltonian is substituted in equation of motion that obtained from the Lagrangian, and the result is classical equations of motion.

\begin{eqnarray}
\frac{1}{\omega_0}\phi_t &=& \delta cos\theta(sec(2g)-cos(2\gamma) tan(2g))+a_0^2cos(2g)(\theta_{xx}csc\theta+\phi_x^2cos\theta) \nonumber\\
\frac{1}{\omega_0}\theta_t &=& \frac{\delta}{2}sin(2\theta) sin(2\gamma) tan(2g)-a_0^2 \phi_{xx}cos(2g)sin\theta \nonumber\\
\frac{1}{\omega_0}g_t &=&- \frac{\delta}{2}sin(2\gamma) sin^2\theta  \nonumber\\
\frac{1}{\omega_0}\gamma_t &=&(4cos(2g)-\delta(cos(2\gamma)(cot(4g)-cos(2\theta) csc(4g))+cos^2\theta sec(2g) )) \nonumber\\
&  &+(cos(2g)(8g_x^2-2\theta_x^2+\frac{1}{2}\phi_x^2(-3+cos(2\theta))-\theta_{xx}cot\theta)+4g_{xx}sin(2g))a_0^2 \nonumber\\
&   &
\end{eqnarray}
These equations describe nonlinear dynamics of non-Heisenberg ferromagnetic chain completely.
Solutions of these equations are magnetic solitons that are not studied in this paper.

In this paper, only the linearized form of equation (10) for small excitation from the ground states is considered. To this end, first, classical ground states must be calculated. therefore in above Hamiltonian only non-derivative part is taken into account:

\begin{eqnarray}
H_0=-J\int \frac{dx}{a_0}(cos^2(2g)+\frac{\delta}{2}(cos^2\theta+sin(2g)cos(2\gamma) sin^2\theta))
\end{eqnarray}
It is necessary to calculate derivative of equation (11) with respect to all variables to find out minimum of $H_0$. As a result, if  $\delta<0$, ground states are at these points:

\begin{eqnarray}
\theta=\frac{\pi}{2},  \gamma=\frac{\pi}{2},&    &{   } sin2g_0=\frac{|\delta|}{4}, | \delta| <4 
\end{eqnarray}

In this paper, only dispersion of spin wave in neighborhood of the ground states is studied.
For this purpose, small linear excitations from the ground states, as shown in eq. (13), are defined:

\begin{eqnarray}
\theta\rightarrow\frac{\pi}{2}-\theta \nonumber\\
2\gamma\rightarrow\pi+\gamma \nonumber\\
2g\rightarrow g_0+g
\end{eqnarray}

In this situation, the linearized classical equations of motion are:

\begin{eqnarray}
\frac{1}{\omega_0}\phi_t &=& \delta (secg_0+tang_0)\theta+a_0^2cosg_0\theta_{xx} \nonumber\\
\frac{1}{\omega_0}\theta_t &=& -a_0^2 \phi_{xx}cosg_0 \nonumber\\
\frac{1}{\omega_0}g_t &=&- \frac{\delta}{2}\gamma  \nonumber\\
\frac{1}{\omega_0}\gamma_t &=&-2(2sing_0+\frac{\delta}{cosg_0})g+4a_0^2 g_{xx}sing_0 \nonumber\\
\end{eqnarray}

Consider functions $\theta$, $\phi$  , $\gamma$   and g  as plane waves to obtain dispersion equation:

\begin{eqnarray}
\phi &=& \phi_0 e^{i(\omega t-kx)}+\bar {\phi_0}e^{-i(\omega t-kx)} \nonumber\\
\theta &=& \theta_0 e^{i(\omega t-kx)}+\bar {\theta_0}e^{-i(\omega t-kx)} \nonumber\\
g &=& g_0 e^{i(\omega t-kx)}+\bar {g_0}e^{-i(\omega t-kx)} \nonumber\\
\gamma &=& \gamma_0 e^{i(\omega t-kx)}+\bar {\gamma_0}e^{-i(\omega t-kx)} \nonumber\\
\end{eqnarray}
 
Substitution of these equations in eq. (14), then:

\begin{eqnarray}
\omega_1^2 &=& \omega_0^2 k^2 a_0^2 (\delta(1+sing_0)+k^2 a_0^2 cos^2g_0) \nonumber\\
\omega_2^2 &=& \omega_0^2 [2sing_0 k^2 a^2_0+\delta (\frac{4\delta}{sin^2g_0}-2sing_0)]
\end{eqnarray}

These equations are dispersion equations of spin wave near the ground states in SU(3) group.

\section{Conclusions}

In this paper, describing equations of one-dimensional anisotropic non-Heisenberg Hamiltonians are obtained using real-parameter coherent states. It was indicated that both dipole and quadruple excitations have different dispersion if there is small linear excitation from the ground state.

In addition, it was indicated that for anisotropic ferromagnets, the magnitude of average quadruple moment is not constant  and its dynamics consists of two parts. One part is rotational dynamics around the classical spin vector ($\gamma_t\ne 0$)  and the other related to change of magnitude of quadruple moment ($g_t\ne 0$).

\end{document}